\documentclass{jaa}
\usepackage{natbib}
\usepackage{graphicx}

\begin{document}\sloppy

\title{Reevaluation of ALMA detection of circumstellar PH$_3$ in the AGB envelope IRC\,+10216: evidence for misidentification with HCN}


\author{M.~Ag\'undez\textsuperscript{1,*}, L. Velilla-Prieto\textsuperscript{1} and J.~P. Fonfr\'ia\textsuperscript{2} and J.~Cernicharo\textsuperscript{1}}
\affilOne{\textsuperscript{1}Instituto de F\'isica Fundamental, CSIC, Calle Serrano 123, E-28006 Madrid, Spain.\\}
\affilTwo{\textsuperscript{2}Observatorio Astron\'omico Nacional, IGN, Calle Alfonso XII 3, E-28014 Madrid, Spain.}


\twocolumn[{

\maketitle

\corres{marcelino.agundez@csic.es}

\msinfo{1 January 2015}{1 January 2015}

\begin{abstract}
The article "Confirmation of interstellar phosphine towards asymptotic giant branch star IRC+10216" by A. Manna and S. Pal uses ALMA data of the C-star envelope IRC\,+10216 to claim a confirmation of the detection of PH$_3$ in this source. The article however incorrectly assign an emission feature observed in the ALMA spectrum of IRC\,+10216 to PH$_3$, while we find that it arises from a highly vibrationally excited state of HCN. Concretely the feature can be confidently assigned to the $J$\,=\,3-2 $\ell=0$ transition of HCN in the $\nu_1$ + 4$\nu_2$ vibrational state based on the observation of the $\ell = +2$ and $\ell = -2$ components of the same rotational transition, $J$\,=\,3-2, with the observed relative intensities in agreement with the relative line strengths. The detection of PH$_3$ in IRC\,+10216 remains confirmed based on the observation of the $J$\,=\,1-0 and $J$\,=\,2-1 lines with the single-dish telescopes IRAM\,30m, ARO SMT\,10m, and \textit{Herschel} \citep{Agundez2008,Agundez2014,Tenenbaum2008}.
\end{abstract}


}]


\doinum{12.3456/s78910-011-012-3}
\artcitid{\#\#\#\#}
\volnum{000}
\year{0000}
\pgrange{1--}
\setcounter{page}{1}
\lp{1}

The molecule phosphine (PH$_3$) was tentatively detected toward the C-rich AGB envelope IRC\,+10216 by \cite{Agundez2008} and \cite{Tenenbaum2008}. These two studies presented the detection of an emission feature located at the frequency of the $J$\,=\,1-0 rotational transition of PH$_3$, 266944.514 MHz, using the IRAM\,30m and the ARO SMT\,10m radiotelescopes, respectively. The feature showed an anomalous line profile composed of a narrow component and a broader one with a U-shape, as observed with the IRAM\,30m telescope, and a flat-topped shape when observed with the ARO SMT\,10m telescope. The narrow component was confidently assigned to the $J$\,=\,15-14 transition of SiS in the $v$\,=\,4 vibrational state by \cite{Agundez2008}, leaving PH$_3$ $J$\,=\,1-0 as the most likely carrier of the broad feature. Confirmation of this assignment came years later with the detection of the $K$\,=\,0 and $K$\,=\,1 components of the $J$\,=\,2-1 rotational transition of PH$_3$, at 533.8 GHz, using the HIFI instrument on board the \textit{Herschel} space telescope \citep{Agundez2014}. The broad nature of the lines of PH$_3$ observed, consistent with the terminal expansion velocity of IRC\,+10216, 14.5 km s$^{-1}$ \citep{Cernicharo2000}, indicated an extended distribution for the emission of PH$_3$ in IRC\,+10216.

\begin{figure*}
\centering
\includegraphics[angle=0,width=\textwidth]{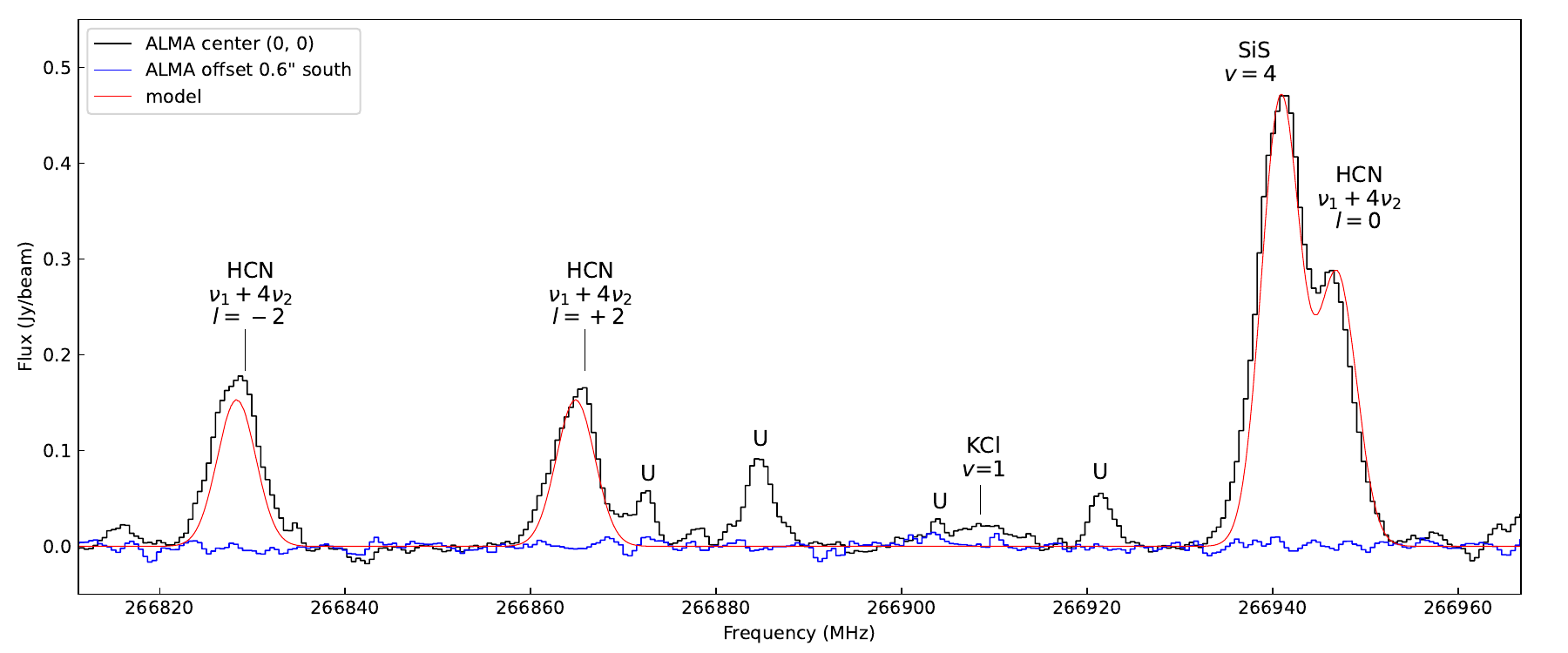}
\caption{ALMA spectrum of IRC\,+10216 at the position of the star (black histogram) and at an offset position located 0.6\,$''$ south from the star. The red line is a simple model in which we assume that the emission comes from a circle of 0.55\,$''$ in diameter with a uniform brightness temperature. The relative intensities of the three HCN lines are set by their relative line strengths.} \label{fig:spectrum}
\end{figure*}

IRC\,+10216 was observed with ALMA with an angular resolution of 0.6\,$''$\,$\times$\,0.5\,$''$ at frequencies covering the PH$_3$ $J$\,=\,1-0 line in project 2011.0.00229.S \citep{Cernicharo2013}. The data from this project analyzed by \cite{Manna2024} shows an emission feature at the frequency of the PH$_3$ $J$\,=\,1-0 line consisting of two Gaussian-like narrow components. The brightest one can be confidently assigned to SiS $v$\,=\,4 $J$\,=\,15-14, as correctly done by \cite{Manna2024}. SiS has been detected in rotational lines from vibrationally excited states up to $v$\,=\,10 \citep{Agundez2012,Velilla-Prieto2015}. Lines from vibrationally excited states are narrow and arise from the hot innermost circumstellar regions close to the AGB star \citep{Velilla-Prieto2023}. The ALMA spectrum shows an additional narrow feature that appears as a shoulder at the high-frequency side of the SiS $v$\,=\,4 $J$\,=\,15-14 line (see Fig.\ref{fig:spectrum} and Figure 1 in \citealt{Manna2024}). This feature is incorrectly assigned to PH$_3$ $J$\,=\,1-0 by these authors. The extended nature of the emission of PH$_3$ in IRC\,+10216, as indicated by the single-dish observations carried out with IRAM\,30, ARO SMT\,10m, and \textit{Herschel}, makes it unlikely that the narrow feature, which arises exclusively from the very inner circumstellar regions, can be assigned to phosphine. In Fig.\,\ref{fig:spectrum} we show the ALMA spectrum at the position of the star and at a offset position located 0.6\,$''$ south of the star. It is clearly seen that the emission of the narrow features disappear at the offset position. The ALMA configuration used during the observations very likely filtered the PH$_3$ emission in IRC\,+10216, which is expected to come from a shell with a diameter of $\sim$\,12\,$''$ \citep{Agundez2014}.

\begin{table}[!b]
\small
\caption{Observed line parameters.}
\label{table:lines}
\centering
\begin{tabular}{lcc}
\hline \hline
Transition & $\nu_{\rm calc}$ & $\nu_{\rm obs} - \nu_{\rm calc}$ \\
& (MHz) & (MHz) \\
\hline
SiS $v$\,=\,4 $J$\,=\,15-14 & 266941.755 & $-$0.842 \\
HCN $\nu_1+4\nu_2$ $J$\,=\,3-2 $\ell=-2$& 266829.233 & $-$0.942 \\
HCN $\nu_1+4\nu_2$ $J$\,=\,3-2 $\ell=+2$& 266865.871 & $-$1.029 \\
HCN $\nu_1+4\nu_2$ $J$\,=\,3-2 $\ell=0$& 266948.781 & $-$1.792 \\
\hline
\end{tabular}
\end{table}

We assign the narrow feature at the high-frequency side of the SiS $v$\,=\,4 $J$\,=\,15-14 line to the $J$\,=\,3-2 $\ell=0$ transition of HCN in the $\nu_1$ + 4$\nu_2$ vibrational state, i.e., ($\nu_1$, $\nu_2^{\ell}$, $\nu_3$) = (1, 4$^0$, 0). HCN has been detected toward IRC\,+10216 in a plethora of vibrationally excited states using the IRAM\,30m telescope \citep{Cernicharo2011} and ALMA \citep{Cernicharo2013}. The spectroscopy of HCN in highly excited vibrational states has been studied by \cite{Mellau2011a,Mellau2011b,Mellau2011c}, which reported the detection of thousands of ro-vibrational lines and provided their frequencies with accuracies better than 0.0007 cm$^{-1}$. We have used the spectroscopic parameters determined in the studies by Mellau to calculate the frequencies of rotational transitions within highly excited vibrational states of HCN. In table\,\ref{table:lines} we list the calculated and observed frequencies of the lines shown in Fig.\,\ref{fig:spectrum}. The difference between calculated and observed frequency for the $J$\,=\,3-2 $\ell=0$ line of HCN in the $\nu_1$ + 4$\nu_2$ state is $-$1.8 MHz, which is within the frequencies errors derived from the data of Mellau, on the order of 1-2 MHz. The spectral resolution of 0.5 MHz of the ALMA spectrum and the fact that line profiles from the very inner regions of the envelope show some asymmetries due to an inhomogeneous emission distribution \citep{Velilla-Prieto2023} result in differences between observed and calculated frequencies which are typically within 1 MHz. The strongest argument in favor of the assignment to HCN comes from the observation of other lines arising from the $\nu_1$ + 4$\nu_2$, state of HCN. The two components with $\ell=\pm2$ are clearly detected at 266829.2 and 266865.9 MHz, with intensities that are fully consistent with the relative line strengths of the $\ell = 0, +2, -2$ components (see Fig.\,\ref{fig:spectrum}). We present in Fig.\,\ref{fig:spectrum} a simple parametric model of the emission of HCN and SiS in which we assume that the emission is distributed as a circle with a diameter of 0.55\,$''$, roughly the size of the synthetic beam, with a uniform brightness temperature. A full model of the emission accounting for the excitation and radiative transfer of HCN and SiS in multiple vibrational states is beyond the scope of this Comment. The main interest of the model spectrum is that it reproduces correctly the relative intensities of the three features that we identify as HCN lines, so that no additional line is necessary to explain the observations. Many other lines of HCN in highly vibrational states are clearly detected in the ALMA spectrum of IRC\,+10216 (see \citealt{Cernicharo2013}). The full census of HCN lines will be presented in detail in a forthcoming publication.

In summary, we have shown that the emission feature assigned to PH$_3$ $J$\,=\,1-0 by \cite{Manna2024} in fact arises from HCN in the $\nu_1$ + 4$\nu_2$ vibrational state. Nonetheless, the detection of PH$_3$ in IRC\,+10216 is solid based on the observation of the $J$\,=\,1-0 and $J$\,=\,2-1 lines with single-dish telescopes \citep{Agundez2008,Agundez2014,Tenenbaum2008}.

\section*{Acknowledgements}

We acknowledge funding support from Spanish Ministerio de Ciencia, Innovaci\'on, y Universidades through grants PID2019-107115GB-C21, PID2019-106110GB-I00, PID2019-105203GB-C21, PID2020-117034RJ-I00, PID2023-147545NB-I00, and PID2023-147545NB-I00. This paper makes use of the following ALMA data: ADS/JAO.ALMA$\#$2011.0.00229.S. ALMA is a partnership of ESO (representing its member states), NSF (USA) and NINS (Japan), together with NRC (Canada), NSTC and ASIAA (Taiwan), and KASI (Republic of Korea), in cooperation with the Republic of Chile. The Joint ALMA Observatory is operated by ESO, AUI/NRAO and NAOJ.

\end{document}